\documentclass[pra,aps,showpacs,floatfix,twocolumn,nofootinbib,superscriptaddress]{revtex4-1}

\usepackage{amsmath}
\usepackage{amsfonts}
\usepackage{amssymb}
\usepackage{graphicx}
\graphicspath{{figures/}}
\usepackage{xcolor} 

\usepackage{physics}
\usepackage{import}
\usepackage{hyperref}  
\usepackage{dcolumn}
\usepackage{bm}
\graphicspath{{figures/}}
\def\redmel#1#2#3{\langle#1\| #2 \| #3 \rangle}

\begin{document}

\preprint{APS/123-QED}

\title{Zero crossings of the differential scalar polarizability of Ba$^+$ clock transition}

\author{N. Jayjong}
\author{M. D. K. Lee}%
\affiliation{Centre for Quantum Technologies, National University of Singapore, 3 Science Drive 2, 117543 Singapore}%
\author{K. J. Arnold}%
\affiliation{Centre for Quantum Technologies, National University of Singapore, 3 Science Drive 2, 117543 Singapore}%
\affiliation{Temasek Laboratories, National University of Singapore, 5A Engineering Drive 1, 117411 Singapore
}%
\author{M. D. Barrett}
 \email{phybmd@nus.edu.sg}
\affiliation{Centre for Quantum Technologies, National University of Singapore, 3 Science Drive 2, 117543 Singapore}%
\affiliation{
Department of Physics, National University of Singapore, 2 Science Drive 3, 117551 Singapore
}%

\begin{abstract}
The differential scalar polarizability $\Delta\alpha_0(\omega)$ of the Ba$^+$ $S_{1/2}-D_{5/2}$ clock transition has a zero crossing near 481nm, which is measured to be  623.603\,13(17)\,THz. From this measurement, we infer a ratio of reduced matrix elements $\redmel{P_{3/2}}{r}{S_{1/2}} /\redmel{P_{1/2}} {r}{S_{1/2}}=1.411\,81(13)$, which provides a stringent test of atomic structure calculations and experimental determination of matrix elements.   Additionally, it enables the construction of an accurate approximation to $\Delta\alpha_0(\omega)$, valid for frequencies up to 450\,THz, with only one reduced matrix element, $\redmel{P_{1/2}}{r}{S_{1/2}}$, appearing in the model's parameterization. We discuss the achievable accuracy of the model, the application to the assessment of blackbody radiation (BBR) shifts in ion-based clocks, and the applicability of the approach to other alkaline-earth ions.
\end{abstract}

\maketitle


\section{Introduction}
\label{sec:Introduction}
The dynamic differential scalar polarizability $\Delta\alpha_0(\omega)$ of a clock transition determines its response to electric fields with the dc value $\Delta\alpha_0(0)$ quantifying the blackbody radiation (BBR) shift \cite{safronova2012blackbody}.   Determination of $\Delta\alpha_0(\omega)$ is typically a significant source of systematic uncertainty in state-of-the-art optical frequency standards \cite{huntemann2016single,brewer2019quantum,hausser2025in+,marshall2025high}.  Recently our group achieved a record-low uncertainty of $1\times10^{-19}$ for the $^{176}$Lu$^+$ $^1S_0-{}^3D_1$ clock transition and demonstrated agreement between two independent systems with a statistically limited uncertainty of $5\times10^{-19}$ \cite{arnold2025optical}.  Despite the low BBR shift for lutetium, uncertainty in $\Delta\alpha_0(0)$ is now the leading systematic and improvement in its determination will enable clock inaccuracies of mid $10^{-20}$ for this system.  

For ions, the determination of $\Delta\alpha_0(0)$ to better than a few percent is challenging.  It typically involves measurements at near-infrared or infrared and extrapolating to dc \cite{huntemann2016single,baynham2018measurement,arnold2018blackbody,arnold2019dynamic}, which is limited by the accuracy of the in situ laser intensity calibration.  Ions having $\Delta\alpha_0(0)<0$ are an exception to this.  In such cases excess micromotion shifts cancel at a so called magic trap drive frequency.  Comparison of a clock with large amounts of EMM against a well-compensated reference system allows the magic trap drive frequency and hence $\Delta\alpha_0(0)$ to be determined with extraordinary precision.  This has been demonstrated in both Sr$^+$ \cite{dube2014high} and Ca$^+$ \cite{huang2019ca} albeit with a significant discrepancy between independent measurements in Sr$^+$ \cite{lindvall2025measurement}.    As proposed in \cite{barrett2019polarizability} and demonstrated in \cite{wei2024improved}, a high accuracy value of $\Delta\alpha_0(0)$ for a reference ion can be transferred to another ion by measuring the ratio of Stark shifts induced by a common laser field.  

In principle, determination of $\Delta\alpha_0(0)$ via a magic trap drive frequency could be used for Ba$^+$ and the $^1S_0-{}^3D_2$ transition in $^{176}$Lu$^+$.  However, the extreme micromotion amplitudes needed to achieve high accuracy makes this approach problematic. The ``magic'' trap drive frequency for Ba$^+$ is $\approx 3.58\,\mathrm{MHz}$, which makes the operating conditions untenable, and large amounts of EMM for Lu$^+$ would likely be correlated with significant ac magnetic field shifts \cite{arnold2024validating,gan2018oscillating}.  Nevertheless, reasonable accuracy of $\Delta\alpha_0(0)$ for Ba$^+$ can be achieved through branching ratio measurements, matrix element determinations, and a zero crossing \cite{barrett2019polarizability}.  This approach also provides a calibration of $\Delta\alpha_0(\omega)$ over a wide wavelength range, consistency checks between multiple precision measurements, and rigorous tests of atomic structure calculations. 

In this work, we determine the zero crossing of $\Delta\alpha_0(0)$ near 481~nm for the $S_{1/2}-D_{5/2}$ transition in Ba$^+$.  This zero crossing is primarily determined by the $S_{1/2}-P_{1/2}$ and $S_{1/2}-P_{3/2}$ transitions at 493\,nm and 455\,nm respectively.  Consequently the zero crossing allows the determination of the ratio $|\langle P_{3/2}\|r\|S_{1/2}\rangle|/|\langle P_{1/2}\|r\|S_{1/2}\rangle|$ which can be compared to independent measurements reported in \cite{woods2010dipole} and theoretical calculations \cite{guet1991relativistic,iskrenova2008theoretical,gopakumar2002electric,safronova2011excitation,porsev2021role,roberts2023electric,cserveny2025theoretical}. It also allows the construction of an accurate model of $\Delta\alpha_0(\omega)$, valid for frequencies up to 450\,THz, with only one reduced matrix element, $\redmel{P_{1/2}}{r}{S_{1/2}}$ appearing in the model's parameterization.

We first provide a summary of the model proposed in \cite{barrett2019polarizability}, the modification that can be made with knowledge of the zero-crossing near 481~nm, and the uncertainties that such modelling introduces.  We then describe the experimental procedure used to determine the zero-crossing.  Using the result, we then provide the determination of the matrix element ratio, an estimation of $\Delta\alpha_0(0)$, and updated recommended values for the $|\langle P_{3/2}\|r\|S_{1/2}\rangle|$, and $|\langle P_{1/2}\|r\|S_{1/2}\rangle|$ reduced matrix elements.  Given the implementation of our approach reported in \cite{wei2024improved}, which used theoretical calculations to extrapolate the polarizability curve for Ca$^+$, we also give an explicit construction of a model for which the extrapolation utilizes published experimental results and much less dependent on theory.  The result is an extrapolation that is almost completely determined by experimental measurements and a factor of three more accurate than that used in \cite{wei2024improved}.

\section{Theory}
As discussed in \cite{barrett2019polarizability}, $\Delta\alpha_0(\omega)$ for the Ba$^+$ clock transition can be approximated by
\begin{multline}
\Delta\alpha_0(\omega)=\frac{c_{614}}{1-(\omega/\omega_{614})^2}-\frac{c_{455}}{1-(\omega/\omega_{455})^2}\\
-\frac{c_{493}}{1-(\omega/\omega_{493})^2}+\frac{c_{0}}{1-(\omega/\omega_{0})^2},
\end{multline}
where the first three terms are the contributions from transitions at the indicated wavelengths, and the last term an approximation to the remaining uv transitions and valence-core corrections with the pole position $\omega_0$ corresponding to an approximate wavelength of 222(5)\,nm as determined from atomic structure calculations.  Frequencies $\omega_{455}$, $\omega_{493}$ and $\omega_{614}$ are taken from \cite{karlsson1999revised}.  Constants $c_k$ determine the strength of each pole and we have
\begin{align}
c_{455}&=\frac{1}{3}\frac{|\langle P_{3/2}\|r\|S_{1/2}\rangle|^2}{\omega_{455}}\\
c_{493}&=\frac{1}{3}\frac{|\langle P_{1/2}\|r\|S_{1/2}\rangle|^2}{\omega_{493}}\\
c_{614}&=\frac{1}{9}\frac{|\langle P_{3/2}\|r\|D_{5/2}\rangle|^2}{\omega_{614}}\nonumber\\
&=\frac{1}{9}\frac{|\langle P_{3/2}\|r\|S_{1/2}\rangle|^2}{\omega_{614}}\left(\frac{\omega_{455}}{\omega_{614}}\right)^3\frac{1-p}{p}\nonumber\\
&=\frac{1}{3}c_{455} \left(\frac{\omega_{455}}{\omega_{614}}\right)^4\frac{1-p}{p}\nonumber\\
&=\frac{1}{3}\mathcal{R} c_{493} \left(\frac{\omega_{455}}{\omega_{614}}\right)^4\frac{1-p}{p}
\end{align}
where $p$ is a measured branching fraction that was reported in \cite{zhang2020branching}, and we have introduced the ratio
\begin{equation}
\mathcal{R}=\frac{c_{455}}{c_{493}}=\frac{|\langle P_{3/2}\|r\|S_{1/2}\rangle|^2}{|\langle P_{1/2}\|r\|S_{1/2}\rangle|^2}\frac{\omega_{493}}{\omega_{455}}.
\end{equation}
With
\[
\mathcal{P}=\frac{1}{3}\left(\frac{\omega_{455}}{\omega_{614}}\right)^4\frac{1-p}{p},
\]
we have
\begin{multline}
\Delta\alpha_0(\omega)=\frac{\mathcal{R} c_{493}}{1-(\omega/\omega_{614})^2}\mathcal{P}-\frac{\mathcal{R}c_{493}}{1-(\omega/\omega_{455})^2}\\
-\frac{c_{493}}{1-(\omega/\omega_{493})^2}+\frac{c_{0}}{1-(\omega/\omega_{0})^2}.
\end{multline}
Using the notation
\[
T^\omega_{\omega_k}=\frac{1-(\omega/\omega_{0})^2}{1-(\omega/\omega_{k})^2},
\]
and the zero crossing near $653\,\mathrm{nm}$ defined by $\Delta\alpha_0(\omega_{653})=0$, gives
\begin{equation}
c_0=\Big[T^{\omega_{653}}_{\omega_{493}}+T^{\omega_{653}}_{\omega_{455}}\mathcal{R}-T^{\omega_{653}}_{\omega_{614}}\mathcal{R}\mathcal{P}\Big]c_{493},
\end{equation}
which can be used to replace the variable $c_0$ with the experimentally measured value of $\omega_{653}$ \cite{chanu2020magic}.  Further, the zero-crossing near 481\,nm, defined by $\Delta\alpha_0(\omega_{481})=0$, is predominately determined by the transitions at $\omega_{455}$ and $\omega_{493}$.  Thus the condition $\Delta\alpha_0(\omega_{481})=0$ leads to an estimate 
\begin{multline}
\label{eq:R}
\mathcal{R}(\omega_{481}, \omega_{653}, \mathcal{P}, \omega_0)\\ =
\frac{
    \left[
        T^{\omega_{481}}_{\omega_{493}}
        - T^{\omega_{653}}_{\omega_{493}}
    \right]
}{
    \left[
        T^{\omega_{481}}_{\omega_{614}}
        - T^{\omega_{653}}_{\omega_{614}} 
    \right] \mathcal{P}
    - 
    \left[
        T^{\omega_{481}}_{\omega_{455}}
        - T^{\omega_{653}}_{\omega_{455}} 
    \right]
},
\end{multline}
which may be used to replace $\mathcal{R}$ with $\omega_{481}$ in the model.

Insofar as the model is accurate, $\mathcal{R}$ can thus be deduced from measured values of $\omega_{481},\omega_{653},\mathcal{P}$ and the estimated value of $\omega_0$.  Similarly $\Delta\alpha_0(\omega)$ can be written
\begin{equation}
\label{eq:Model}
\alpha_0(\omega)=\mathcal{C}(\omega,\omega_{481}, \omega_{653}, \mathcal{P}, \omega_0) \, c_{493}
\end{equation}
where $c_{493}$ is determined by $\redmel{P_{1/2}}{r}{S_{1/2}}$, which can be accurately measured \cite{woods2010dipole}, and the function $\mathcal{C}$ is determined by two measured zero crossings, a single measured branching fraction and weakly dependant on a theoretical estimate of the approximating uv pole position.

To investigate the validity Eq.~\ref{eq:R} and Eq.~\ref{eq:Model} and the predictive power they have for $\mathcal{R}$ and $\Delta\alpha_0(\omega)$, we first use the atomic structure calculations given in \cite{barrett2019polarizability} as a possible instance describing the real atom. Using the parameters $\omega_{481},\omega_{653},\mathcal{P}, c_{493}$ and $\omega_0$ as determined from the given instance, we compare the values of $\mathcal{R}$ and $\Delta\alpha_0(\omega)$ determined from Eq.~\ref{eq:R} and Eq.~\ref{eq:Model} to those obtained from the instance.  For any given instance, $\omega_{481},\omega_{653},\mathcal{P}$ and $c_{493}$ are well-defined parameters, and only $\omega_0$ is subject to interpretation. We take $\omega_0$ that best represents $\Delta\alpha_0(\omega)$ through Eq.~\ref{eq:Model}.  

\begin{figure}
\begin{center}
  \includegraphics[width=0.9\linewidth]{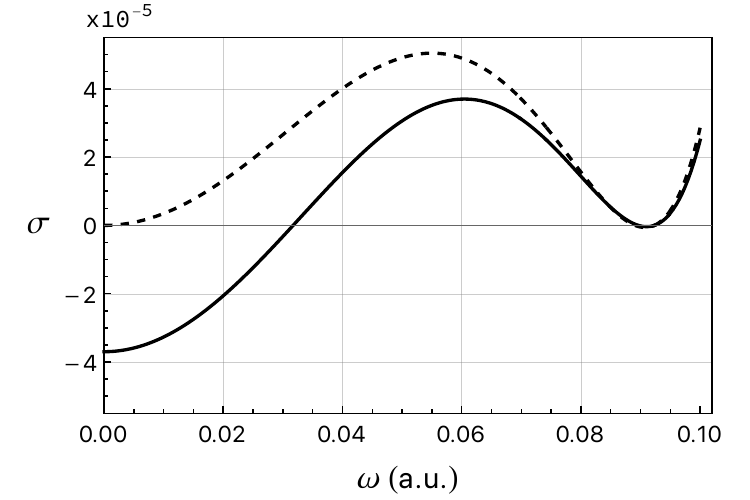}
  \caption{Fractional error due to the use of Eq.~\ref{eq:Model} to represent $\Delta\alpha_0(\omega)$, as determined from atomic structure calculations.  Solid: $\omega_0\approx 0.204\,580\,\mathrm{a.u.}$ (222.717\,nm) minimizes the maximum absolute fractional error for wavelengths above 700\,nm ($\omega \approx 0.065\,\mathrm{a.u.}$).  Dashed: $\omega_0\approx 0.204\,696\,\mathrm{a.u.}$ (222.590\,nm) gives agreement at $\omega=0$ between the model and the full calculation.} 
  \label{ModelErrors}
\end{center}
\end{figure}

Choosing $\omega_0\approx 0.204\,580\,\mathrm{a.u.}$ (222.717\,nm) minimizes the maximum absolute fractional error in $\Delta\alpha_0(\omega)$ for $\omega \lesssim 0.065\,\mathrm{a.u.}$ ($\lambda\gtrsim700\,\mathrm{nm}$) and gives a fractional error in $\mathcal{R}$ of $-2.5 \times 10^{-5}$.  The former is plotted in Fig.~\ref{ModelErrors} over the range $\omega<0.1\,\mathrm{a.u.}$ $(\lambda\gtrsim 450\,\mathrm{nm})$.  For comparison, we also give the result for the choice $\omega_0\approx 0.204\,696\,\mathrm{a.u.}$ (222.590\,nm), which gives agreement in $\Delta\alpha_0(0)$ and a fractional error in $\mathcal{R}$ of $-2.9 \times 10^{-5}$.  Although Eq.~\ref{eq:Model} remains accurate for wavelengths in the extended range $\gtrsim 450\,\mathrm{nm}$, zero-crossings and resonant transitions that occur below 700\,nm give rise to large values and variations in $\Delta\alpha_0(\omega)$.  Hence we are primarily interested in the region $\omega<0.065\,\mathrm{a.u.}$ ($\lambda\gtrsim700\,\mathrm{nm})$ where $\Delta\alpha_0(\omega)$ is relatively flat.

We emphasise that the errors given for $\mathcal{R}$ and $\Delta\alpha_0(\omega)$ are estimates in modelling errors that arise from the uv pole approximation and not to be confused with uncertainties that arise from parameter estimation, which will typically be much larger.  The real atom will undoubtedly yield different parameters from those determined from atomic structure calculations, but that also changes the model.  Thus it is prudent to consider the accuracy of the model over a wider class of representations for the atom, and not just a single instance.  

Contributions to $\Delta\alpha_0(\omega)$ from the three transitions at 614, 493, and 455\,nm are heavily constrained by experimental results, so the only difference in the model and the real atom lies in approximating all uv terms by a single transition.  As there is no significant contributions from uv transitions out of the ground state, there is no significant cancellation of terms in a Taylor expansion of the uv terms, which can substantially change the nature of a single pole approximation or even its validity.  Thus a single pole approximation accurate to second order in ($\omega/\omega_0$) is guaranteed.  Degradation of the approximation only arises from combining transitions at distinct frequencies.   

Approximately 80\% of the uv contributions arise from $D_{5/2}$ transitions to the $F_{5/2}$ and $F_{7/2}$ doublet of the $4f$ electron configuration.  Both transitions are very close to 234\,nm and the $F_{7/2}$ contribution is 20 times larger than that of $F_{5/2}$, which is consistent with LS coupling.  Thus degradation of the model arises only from the remaining 20\%.  Half of the remainder arises from $D_{5/2}$ transitions to $nf$ configurations with $n>7$.  In calculations this is represented as a single pole with a transition frequency $\omega=0.3030\,\mathrm{a.u.}$ ($\lambda=150.386\,\mathrm{nm}$), which is given by the $n=8$ transition frequency.  We will henceforth label this as the tail.

Taking only the $4f$ and tail contributions as a representation of the true uv terms leads to practically identical modelling errors as found using the full calculation illustrating the negligible impact that all other uv terms have on the accuracy of Eq.~\ref{eq:R} and Eq.~\ref{eq:Model}.  Thus it is sufficient to consider modelling errors using only the $4f$ and tail contributions.  Since the $4f$ terms are expected to be reasonably well estimated, we vary only the parameters of the tail represented by $c_1/(1-(\omega/\omega_1)^2)$.  For $0<c_1<4$ and $\omega_1>0.2\,\mathrm{a.u.}$
\begin{equation}
\frac{\delta \mathcal{R}}{\mathcal{R}}\approx 0.66 \frac{\delta\Delta\alpha_0(0)}{\Delta\alpha_0(0)}\approx \beta(\omega_1)c_1
\end{equation}
are fairly well satisfied.  The proportionality constant $\beta$ is only weakly dependent on $c_1$ and is shown as a function of $\omega_1$ alone.  As $\omega_1$ approaches the pole positions of the $4f$ contributions $\beta\rightarrow 0$ as expected.  For large $\omega_1$ it limits to approximately $-3.4\times 10^{-5}$ and is slightly less than half of that at the ionization limit of the $D_{5/2}$ level ($\omega=0.3417\,\mathrm{a.u.}$).  For $c_1\in(0,4)\,\mathrm{a.u.}$ and $\omega_1\in(0.3030,0.3417)\,\mathrm{a.u.}$, which allows for frequencies up to the ionization threshold and up to a factor of two increase in the tail's estimated contribution, fractional errors in  $\mathcal{R}$ and $\Delta\alpha_0(0)$ are bounded by $-5.8\times 10^{-5}$ and $-8.8\times10^{-5}$ respectively.  Since larger frequencies will more significantly degrade the single pole representation, this simplified analysis will be a more conservative bound than taking random variates of all contributing terms of a full calculation.  So long as uncertainties from parameter estimates do not approach these limits, modelling errors can be safely ignored.

\section{Measurement of the zero crossing}
Following the approach given in \cite{chanu2020magic}, the zero-crossing of $\Delta\alpha_0(\omega)$ is found using a linearly polarized laser near 481\,nm focussed onto the ion and measuring the induced ac-Stark shift of the clock transition as a function of the laser frequency.  The shift is given by \cite{le2013dynamical}
\begin{multline}
\delta_s(M_J)=-\frac{1}{2}\Delta\alpha_0(\omega)\langle E^2\rangle\\
-\frac{1}{4}\alpha_2(\omega)\frac{3 M_J^2-J(J+1)}{J(2J-1)}\left(3\cos^2\theta-1\right)\langle E^2\rangle,
\end{multline}
where $M_J$ denotes the applicable eigenstate of $D_{5/2}$, $\alpha_2(\omega)$ is the dynamic tensor polarizability of $D_{5/2}$, $\theta$ is angle between the 481-nm laser polarization and the quantization axis, and $\langle E^2\rangle$ is the time-averaged square of the laser electric field, which is proportional to the laser intensity.  The scalar and tensor components, $\delta_0$ and $\delta_2$ respectively, can be determined by the orthogonal combinations
\begin{align}
\delta_0(\omega)&=\frac{1}{3}\left[\delta_s(1/2)+\delta_s(3/2)+\delta_s(5/2)\right]\nonumber\\
&=-\frac{1}{2} \Delta\alpha_0(\omega)\langle E^2\rangle\\
\intertext{and}
\delta_2(\omega)&=\frac{25}{42}\left(\delta_s(5/2)-\frac{1}{5}\delta_s(3/2)-\frac{4}{5}\delta_s(1/2)\right)\nonumber\\
&=-\frac{1}{4}\alpha_2(\omega)\left(3\cos^2\theta-1\right)\langle E^2\rangle.
\end{align}
The ratio
\begin{equation}
\frac{\delta_0}{\delta_2}=\frac{\Delta \alpha_0(\omega)}{\alpha_2(\omega)(3\cos^2\theta-1)},
\end{equation}
is then independent of slow variations in the laser intensity but has the same zero-crossing.

\begin{figure*}[t] 
\begin{center}
        \includegraphics{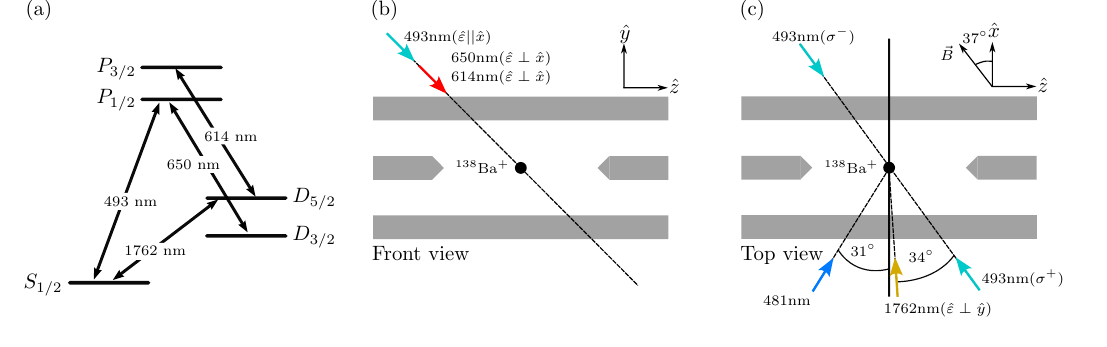}
    \caption{%
    (a) Relevant energy levels and transitions.  
    (b) Front view of the trap showing the orientation of the Doppler cooling and detection lasers at 493 and 650~nm, and the repumping laser at 614~nm.  
    (c) Top view of the trap showing the orientation of the clock laser at 1762\,nm, which is aligned 34$^\circ$ to the magnetic field to allow driving the $|S_{1/2}, \pm 1/2\rangle \to |D_{5/2}, \pm M_J\rangle$ transitions for $M_J = 1/2, 3/2, 5/2$ efficiently; the 493\,nm $\sigma^{\pm}$ beams, which are used for optical pumping into $|S_{1/2}, \pm 1/2\rangle$;  and the Stark shifting laser at 481\,nm.  The magnetic field is aligned along the propagation directions of the 493\,nm $\sigma^{\pm}$ beams. Linear polarization of the 481\,nm laser lies either in the $xz$-plane (Config.~I) or along the $y$-axis (Config.~II).
    }
    \label{fig:setup}
\end{center}
\end{figure*}

Measurements are carried out in a linear Paul trap with axial end-caps as described elsewhere \cite{zhao2025lande,lee2026precision}.  The relevant level structure of $^{138}$Ba$^+$ is illustrated in Fig.~\ref{fig:setup}(a). Doppler cooling is provided by driving the 493- and 650-nm transitions, with light scattered at 650\,nm collected onto a single photon counting module (SPCM) for detection.  The $D_{5/2}$ level is populated by driving the $S_{1/2}-D_{5/2}$ clock transition at 1762\,nm and depopulated by driving the $D_{5/2}-P_{3/2}$ transition at 614\,nm.  State preparation into the $m=\pm1/2$ states of the $S_{1/2}$ levels is facilitated by two additional 493-nm beams that are $\sigma^\pm$ polarized.

Laser configurations are similar to that used in  \cite{chanu2020magic} and illustrated in Fig.~\ref{fig:setup}(b) and (c). Cooling and repumping light at 493, 614, and 650\,nm are all collinear and propagate at 45 degrees to the $z$-axis.  All three beams are linearly polarized with the  650-nm and 614-nm lasers polarized orthogonal to the 493-nm laser, which is polarized along the imaging axis ($\hat{\mathbf{x}}$).  The 493\,nm $\sigma^\pm$ beams lie in the horizontal ($xz$) plane at an angle of approximately 37 degrees to the $x$-axis.  They are aligned collinear with a $0.1\,\mathrm{mT}$ magnetic field, which is sufficient to lift the Zeeman degeneracy and prevent dark states when driving the $D_{3/2}-P_{1/2}$ and $D_{5/2}-P_{3/2}$ transitions.  A Stark-shifting laser near 481\,nm propagates in the $xz$ plane at an angle of approximately 31 degrees to the $x$-axis.  The angle is limited by available optical access and of no particular consequence.  The beam is linearly polarized either in the $xz$-plane (Config. I) or orthogonal to the magnetic field (Config. II). All lasers are switched by acousto-optic modulators (AOMs).  Although these typically have extinction ratios above 40\,dB, measurements are made differentially, i.e. with and without the Stark-shifting beam, so AOM performance is not critical.

The 481\,nm laser is generated via frequency doubling in a bow-tie cavity. The fundamental 962\,nm light is produced by a tunable external-cavity diode laser (ECDL) and amplified using a tapered amplifier (Coherent, TA-0950-2000).  Its wavelength was measured with a HighFinesse WS8-10 wavemeter with an accuracy of $\sim10\,\mathrm{MHz}$ (corresponding to 20\,MHz at 481\,nm). The 962\,nm laser was locked to the bow-tie cavity. The long-term drift of the cavity was stabilized by frequency-locking the cavity to the wavemeter, which was periodically calibrated using a cesium-stabilized 852\,nm laser. The 481\,nm beam power was monitored after passing through the chamber via a photodetector and power stabilized using a double-pass AOM.

As described in \cite{arnold2020precision}, the clock laser at 1762\,nm is phase-locked to a frequency comb, which is locked to an ultra-stable reference cavity.  It propagates almost co-linear with the $x$-axis and is linearly polarized in the horizontal plane, which permits efficient coupling to the $\ket{S_{1/2},\pm 1/2}-\ket{D_{5/2},\pm M_J}$ transitions for $M_J=1/2,3/2,$ and $5/2$.  The laser is continuously on with a detuning of approximately -325\,MHz from resonance. The transition of interest is driven by the first order sideband of a wideband electro-optic modulator (EOM), which is frequency shifted near to resonance for clock interrogation.  In this way the total laser intensity, and corresponding ac-Stark shift, is constant for all transitions.  Control of the radio frequency (rf) power driving the EOM enables equal $\pi$-times for each $M_J$ transition.   This scheme of keeping the rf-drive on and only switching the frequency is necessary to suppress phase chirp effects.

A typical experiment consists of four steps:  $300\,\mathrm{\mu s}$ of Doppler cooling, which includes the 614-nm beam to facilitate repumping from $D_{5/2}$; optical pumping for $100\,\mathrm{\mu s}$ into $\ket{S_{1/2},M_J=\pm1/2}$, which includes the 650-nm beam to counter any pumping into $D_{3/2}$; a $300\,\mathrm{\mu s}$ Rabi clock interrogation to $\ket{D_{5/2},\pm M_J}$, and finally detection.  State detection uses an adapative Bayes method \cite{myerson2008high} with non-deterministic duration but typically takes 100$\,\mathrm{\mu s}$.  When including the 481-nm laser to Stark shift the transition, the beam is switched on $10\,\mathrm{\mu s}$ before the clock laser to allow the active laser intensity to settle.

\begin{figure*}[!ht]
  \centering
    \includegraphics[width=0.48\linewidth]{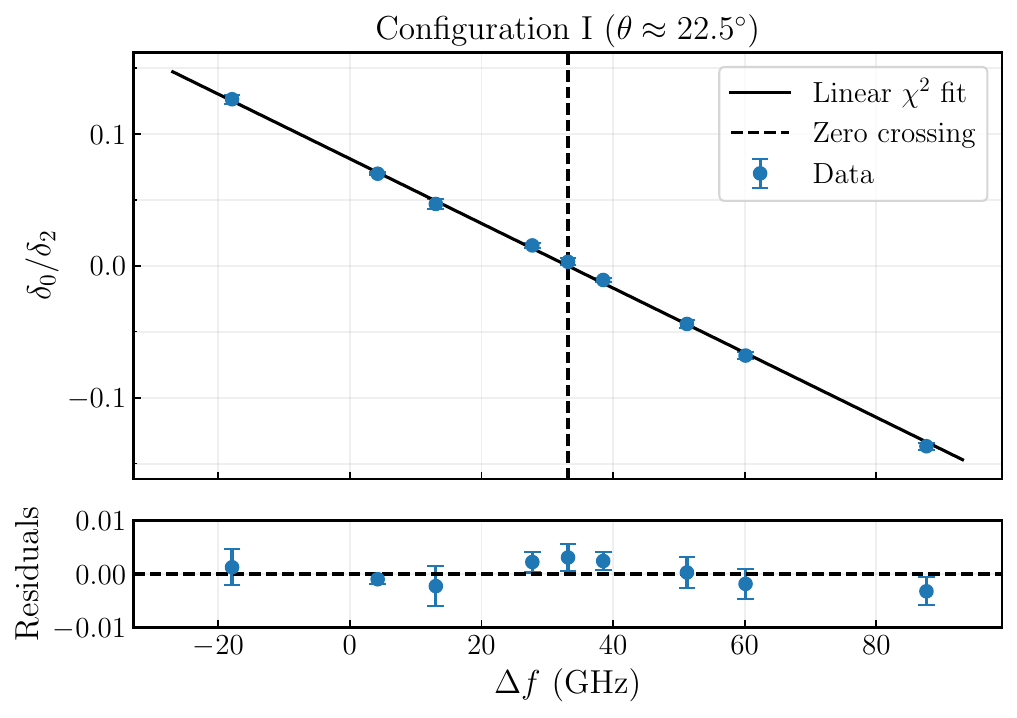}\hspace{0.5cm}
    \includegraphics[width=0.48\linewidth]{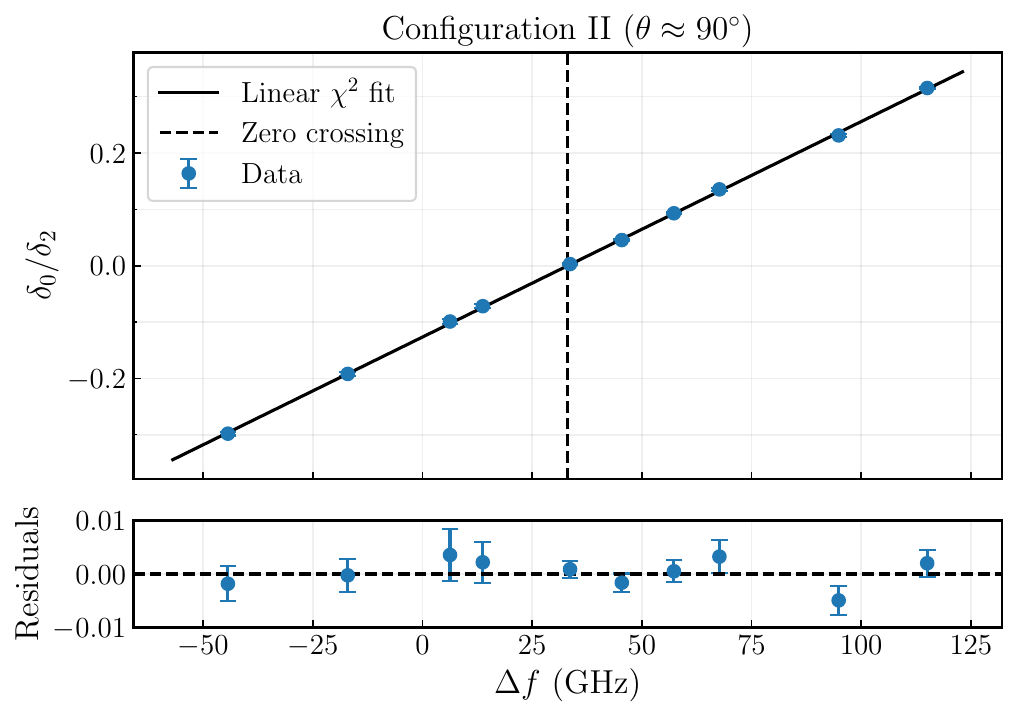}
  \caption{Ratio of scalar to tensor shifts $\delta_0(\omega)/\delta_2(\omega)$ as a function of frequency detuning relative to 623.570\,THz for two polarization configurations of 481 nm. The black solid lines indicate linear $\chi^2$ fits, and the vertical black dashed lines indicate the zero crossings obtained from the fits.}
    \label{data}
\end{figure*}

For each $M_J=1/2, 3/2,$ and $5/2$, the clock transition is realized by steering the clock laser to the average of the Zeeman pair $\ket{S_{1/2},\pm 1/2}-\ket{D_{5/2},\pm M_J}$ to cancel linear Zeeman shifts of both levels.  Using an interleaved servo in which the optical transition is alternately interrogated both with and without the 481-nm laser, the shift is determined from the difference frequency in the two configurations.  In a single clock cycle, all three Zeeman pairs are measured: each of the six transitions $\ket{S_{1/2},\pm1/2} - \ket{D_{5/2},\pm M_J}$ for $M_J=1/2, 3/2,$ and $5/2$ is interrogated both sides of each transition, both with and without the 481-nm Stark shift beam.  Each cycle of 24 measurements is repeated $N=100$ times.  From these measurements error signals are derived to track the center and splitting for each Zeeman pair $\ket{S_{1/2},\pm1/2} - \ket{D_{5/2},\pm M_J}$, both with and without the Stark shift.  The ratio $\delta_0(\omega)/\delta_2(\omega)$ is inferred at each step from the frequency differences.  The servo is first run for several clock cycles, to ensure each shifted and unshifted Zeeman pair is being tracked, and $\delta_0(\omega)/\delta_2(\omega)$ is inferred from a further 500 updates.  The measurements were performed at nine (ten) different frequencies in Config.~I (Config. II) spanning a range of approximately 100\,GHz (150\,GHz), and the results are shown in Fig.~\ref{data}.

The zero-crossing for each configuration was determined by linear regression.  We obtain a zero-crossing of $\omega_{481}=2\pi\times 623.603\,15(28)\,\mathrm{THz}$ ($\chi^2_\nu=1.04$) for Config.~I and $\omega_{481}=2\pi\times623.603\,11(21)\,\mathrm{THz}$ ($\chi^2_\nu=0.83$) for Config.~II, giving a weighted mean of $\omega_{481}=2\pi\times623.603\,13(17)\,\mathrm{THz}$.  As the slopes are related we also infer the angle $\theta$ for Config.~I to be $22.50(49)^\circ$ and the slope of $2\Delta\alpha_0(\omega)/\alpha_2(\omega)$ near the zero crossing to be $m=-3.824(20)/\,\mathrm{THz}$.  These values can also be estimated using a joint fit to both data sets, which gives practically identical results as expected from the agreement in the independently estimated zero-crossings.

As measurements are done differentially, systematics are heavily suppressed.  However, near-resonant cooling light can give rise to very large Stark shifts, which may give pause for concern.  For our typical operating parameters, the Stark shift from the cooling light with the AOM on would be $\approx 1\,\mathrm{MHz}$, which is consistent with the $\approx 10^{6}\,\mathrm{s}^{-1}$ measured scattering rate out of $S_{1/2}$ with this beam.  From the slope of the polarizability curve at the zero crossing, the 0.17\,GHz uncertainty would correspond to a 0.4\,Hz differential shift from the Stark shift beam.  Thus, for the finite extinction ratio of the cooling light to cause a problem, a similar differential shift would need to arise.  Leakage light from the double pass AOM switching the cooling light is unshifted so the light is further detuned. This has been confirmed both by disconnecting the rf source from the AOM, and by measuring the scattering rate out of $S_{1/2}$ from the cooling light with the AOM switched off.  From the latter measurement, we infer a residual Stark shift from the cooling light below $1\,\mathrm{Hz}$.  This will be substantially suppressed as a common mode shift due to the differential nature of the measurement.  Optical pumping beams for state preparation have much lower intensity, and hence of much less concern.

\section{Estimation of $\mathcal{R}$ and $\Delta \alpha_0(\omega)$}
From the measured zero crossing, and parameters 
\begin{align*}
\omega_{653}&=2\pi \times 459.1614(28)\,\mathrm{THz},\\
 p&=0.763\,107(65),\\
 \omega_0&= 2 \pi\times 1\,350(30)\,\mathrm{THz},
\end{align*}
from \cite{chanu2020magic}, \cite{zhang2020branching}, and \cite{barrett2019polarizability} respectively, we estimate
\begin{equation}
\mathcal{R}=1.839\,68(32)
\end{equation}
or equivalently
\begin{equation}
R_0=\frac{\redmel{P_{3/2}}{r}{S_{1/2}}}{\redmel{P_{1/2}}{r}{S_{1/2}}}=1.411\,81(13).
\end{equation}
Uncertainties from the zero-crossings contribute negligibly to the total uncertainty.  Fractionally, the uncertainty contribution from $\omega_0$ is approximately $\sqrt{2}$ larger than that from $\mathcal{P}$.  When combined the two contributions make up $\approx 99\%$ of the total fractional uncertainty for either ratio.   Any modelling bias or uncertainty has been neglected, as they are a fraction of the total uncertainty.  

Reduced matrix elements $\redmel{P_{1/2}}{r}{S_{1/2}}$ and $\redmel{P_{3/2}}{r}{S_{1/2}}$ given in \cite{woods2010dipole} were determined using resonant excitation Stark ionization spectroscopy (RESIS), in which spectroscopy of non-penetrating high-$L$ Rydberg states is used to infer the properties of the ionic core.  Fine structure patterns lead to a determination of the dipole polarizability of the ion core.  For a Ba$^+$ core this is given by
\begin{multline}
\label{Eq:alpha0}
\alpha_0=\frac{1}{3}\left[\frac{|\langle P_{1/2}\|r\|S_{1/2}\rangle|^2}{\omega_{493}}+\frac{|\langle P_{3/2}\|r\|S_{1/2}\rangle|^2}{\omega_{455}}\right]\\
+\alpha_\mathrm{core}+\alpha_\mathrm{vc}+\alpha_\mathrm{tail},
\end{multline}
where $\alpha_\mathrm{core}$ in this context is the polarizability of Ba$^{2+}$.  Measurements of $\alpha_0$ and $\alpha_\mathrm{core}$ were given in \cite{woods2009dipole} and \cite{woods2010dipole} respectively, and $\alpha_\mathrm{vc}$ and $\alpha_\mathrm{tail}$ are small corrections from theory.  Contributions to $K$ splittings of individual Rydberg levels are also dominated by the reduced matrix elements of interest but their contributions tend to cancel \cite[Eq. 2]{woods2010dipole}.  Thus the combination of measurements allows the individual matrix elements to be determined.  

\begin{figure}[t]
  \centering
    \includegraphics[width=0.9\linewidth]{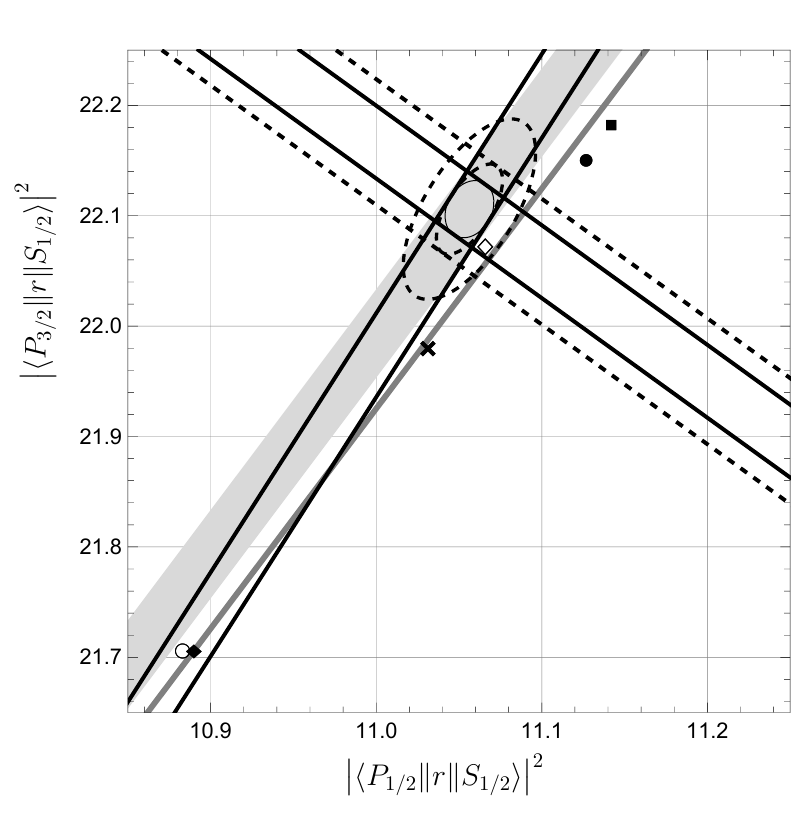}\hspace{0.5cm}
  \caption{Summary of the measurements given in \cite{woods2010dipole} and here.  Solid black lines are as given in \cite[Fig. 4]{woods2010dipole} with the inner ellipse giving the $1\sigma$ uncertainty.  The dotted lines include the additional uncertainty from theory contributions and we give the corresponding $1\sigma$ and $2\sigma$ uncertainty boundaries.  Boundaries are calculated assuming there is no correlation in the uncertainties defining the straight lines.  The light grey region gives the ratio $R_0^2$ given \cite{woods2010dipole} and the dark grey region the ratio reported here with the width in both cases indicating the $1\sigma$ uncertainty.  Single points denote theoretical values from  \cite{guet1991relativistic} (\resizebox{5.5pt}{4.5pt}{$\blacklozenge$}), \cite{iskrenova2008theoretical} (\resizebox{5pt}{5pt}{$\bullet$}), \cite{gopakumar2002electric} (\resizebox{5pt}{5pt}{$\diamond$}), \cite{safronova2011excitation} (\resizebox{4pt}{4pt}{$\blacksquare$}), \cite{porsev2021role} (\resizebox{5pt}{5pt}{$\circ$}), and \cite{roberts2023electric,cserveny2025theoretical}($\boldsymbol{\pmb{\times}}$).}
  \label{Measurements}
\end{figure}

In essence our measurement of the ratio plays the same role as the use of $K$ splittings in \cite{woods2010dipole}.  The zero crossing at $\omega_{481}$ is dominated by contribution from the two transitions of interest, which tend to cancel.  The contribution from the 614\,nm transition is pinned to the $P_{3/2}$ contribution by the branching ratio and contributions from uv transitions are pinned by the zero crossing near 653\,nm insofar as their frequency dependence is captured by $\omega_0$.   We can therefore rewrite Eq.~\ref{Eq:alpha0} in the form
\begin{equation}
|\langle P_{1/2}\|r\|S_{1/2}\rangle|^2=\frac{3\omega_{493}}{1+\mathcal{R}}(\alpha_0-\alpha_\mathrm{core}-\alpha_\mathrm{vc}-\alpha_\mathrm{tail})
\end{equation}
We take $\alpha_0=123.88(5)$ \cite{woods2009dipole},  $\alpha_\mathrm{core}=10.75(10)$ \cite{woods2010dipole}, $\alpha_\mathrm{vc}=-0.51(14)$ \cite{iskrenova2008theoretical,barrett2019polarizability}, and $\alpha_\mathrm{tail}=0.064(64)$.  Here $\alpha_\mathrm{tail}$ refers to all other contributing transitions, which are taken from \cite{barrett2019polarizability}.  They are assumed to have 100\% uncertainty with worst possible correlation, although this has negligible impact on final uncertainties.  We then obtain
\begin{subequations}
\begin{align}
\label{Eq:MErevised}
\redmel{P_{1/2}}{r}{S_{1/2}}&=3.3282(28),\\
\redmel{P_{3/2}}{r}{S_{1/2}}&=4.6988(39),
\end{align}
\label{Eq:MErevisedA}
\end{subequations}
for the reduced matrix elements.  Although these are consistent with those given \cite{woods2010dipole}, there are two key differences.  Firstly, the uncertainties for the two matrix elements are strongly correlated due to the tight bounds on the ratio $\mathcal{R}$ (or $R_0$) we give here.  Secondly, the uncertainties are slightly larger despite the tighter constraint on the ratio.  The constraints shown in \cite[Fig. 4]{woods2010dipole}, which are defined by Eq.~\ref{Eq:alpha0}, do not appear to have included any uncertainty in the correction terms $\alpha_\mathrm{vc}$ or $\alpha_\mathrm{tail}$, but the uncertainty from $\alpha_\mathrm{vc}$ is the largest contribution.  Thus, the uncertainties given in \cite{woods2010dipole} will be under-represented.  However, the uncertainty for the ratio would not significantly change as it is most strongly determined by the $K$-splittings.  A summary of these considerations is illustrated in Fig.~\ref{Measurements} where we also indicate theoretical values taken from \cite{guet1991relativistic,iskrenova2008theoretical,gopakumar2002electric,safronova2011excitation,porsev2021role,roberts2023electric,cserveny2025theoretical}.  We note that the theoretical value of $1.4116(2)$ given in \cite{roberts2023electric,cserveny2025theoretical} agrees well with our measured value.   In addition their reported matrix elements are within $1.5\sigma$ of the inferred values given in Eq.~\ref{Eq:MErevisedA}.

\begin{figure}[!ht]
  \includegraphics[width=0.9\linewidth]{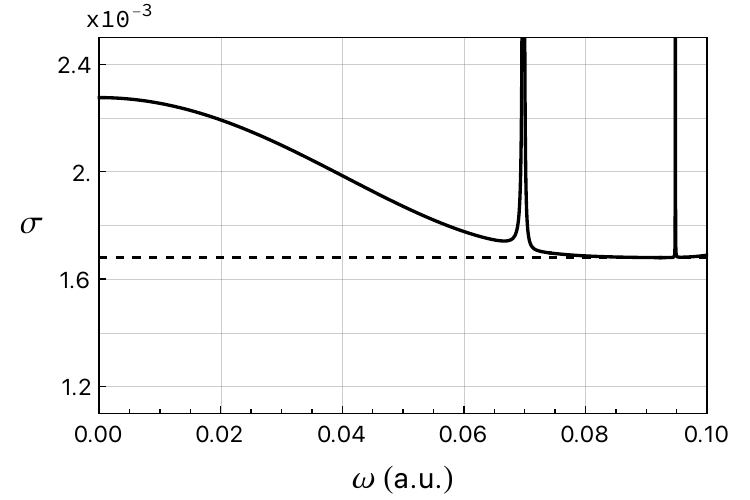}
  \caption{Total fractional uncertainty in $\Delta\alpha_0(\omega)$.  The dashed line shows the contribution arising from $c_{493}$.  The remainder is primarily due to the uncertainty in $\omega_0$.  Uncertainties from $\omega_{481}$ presented here and $\omega_{653}$ reported in \cite{chanu2020magic} result in the divergences at those points with negligible contribution otherwise.}
  \label{fig:DSP}
\end{figure}

The matrix element given in Eq.~\ref{Eq:MErevised} can be used to determine $\Delta\alpha_0(\omega)$ over a wide frequency range.  In Fig.~\ref{fig:DSP} we plot the fractional uncertainty in the estimation of $\Delta\alpha_0(\omega)$ over a frequency range corresponding to wavelengths $\gtrsim450\,\mathrm{nm}$.  Over this frequency range the uncertainty is almost completely determined by the uncertainty in the matrix element and the determination of $\omega_0$, with the latter having most influence at $\omega\approx 0$.  Uncertainties in the zero crossings are only significant near the zero-crossings, which results in divergences at those points.  Of most interest is $\Delta\alpha_0(0)$ for which we obtain
\begin{equation}
\Delta\alpha_0(0)=-73.33(17)\,\mathrm{a.u.}
\end{equation}
The value of $\Delta\alpha_0(0)=-73.56(21)\,\mathrm{a.u.}$ given in \cite{chanu2020magic} was derived using both matrix elements reported in \cite{woods2010dipole}.  The slight difference is due to the difference in $R_0$ reported here with that given in \cite{woods2010dipole}.

\section{Application to other systems}
As was noted in \cite{barrett2019polarizability}, characterization of $\Delta\alpha_0(\omega)$ as used here could equally be done for other alkaline-earth ions such as Sr$^+$, Ca$^+$ and Ra$^+$.  Application to Ca$^+$ was specifically noted as all required measurements were already reported in the literature, and it was of relevance to Al$^+$ clocks using a co-trapped Ca$^+$ ion for quantum logic spectroscopy.  This approach has now been reported in \cite{wei2024improved}, but the authors used theoretical calculations to extrapolate the polarizability curve to the measurement wavelength.  For comparison, we give an extrapolation using published experimental results, which practically eliminates the dependence on theory and provides a more accurate result.

Following \cite{barrett2019polarizability}, $\Delta\alpha_0(\omega)$ for the Ca$^+$ clock transition maybe written
\begin{multline}
\label{CaDSP}
\Delta\alpha_0(\omega)=\Delta\alpha_0(0)+\tfrac{2 M^2 \bigl(\tfrac{\omega_{393}}{\omega_{854}}\bigr)^{\!3} \tfrac{p_1}{p_2}}{9\omega_{854}}f\bigl(\tfrac{\omega}{\omega_{854}}\bigr)\\
-\tfrac{2 M^2 }{3\omega_{393}}f\big(\tfrac{\omega}{\omega_{393}}\big)-\tfrac{M^2 }{3\omega_{397}}f\bigl(\tfrac{\omega}{\omega_{397}}\bigr)\\
+\sum_k c_k f(\omega/\omega_k)
\end{multline}
where $f(x)=x^2/(1-x^2)$, $\omega_x$ denotes the frequency of the contributing transition at the indicated approximate wavelength (in nm), $\Delta\alpha_0(0)=-44.079(13)$ is the dc value reported in \cite{huang2019ca}, $p_1=0.0587(2)$ and $p_2=0.9347(3)$ are the branching fractions reported in \cite{gerritsma2008precision}, and $M=2.8928(43)$ is the reduced matrix element $\langle P_{1/2}\|r\| S_{1/2}\rangle$ reported in \cite{hettrich2015measurement}.  The expression assumes $\langle P_{3/2}\|r\| S_{1/2}\rangle/\langle P_{1/2}\|r\| S_{1/2}\rangle=\sqrt{2}$, which is expected to be a very good approximation for Ca$^+$\cite{safronova2011blackbody,roberts2023electric}.  The summation is over all contributing uv transitions.  As noted in \cite{safronova2011blackbody}, the uv terms in Ca$^+$ provide a significant tail that converges slowly making it difficult to calculate.  However, here the dc component is factored out and included in the experimental value of $\Delta\alpha_0(0)$.  Hence, the influence of the tail is heavily suppressed by the frequency dependence, as the uv transitions are even deeper in the uv than in Ba$^+$.

As with Ba$^+$, the dominant uv term arises from the $D_{5/2}$ to $4F_{7/2}$ transition, which would give a correction of $2.39(5)f(\omega/\omega_{184})$ \cite{safronova2011blackbody}.  Including transitions up to $7F_{7/2}$, the correction can be written $3.57f(\omega/\omega_{175.7})$, which is correct to $4^\mathrm{th}$ order.  Other contributions listed in \cite{safronova2011blackbody} are lumped together in groups.  If these are included with the longest possible wavelength for each group given to the frequency dependence, which would tend to overestimate the contribution, we obtain $4.71f(\omega/\omega_{167.3})$.  We use $4.71f(\omega/\omega_{167.3})$ to estimate the correction and the difference from $2.39f(\omega/\omega_{184})$ as an indication of uncertainty.

With exception of the summation term, Eq.~\ref{CaDSP} is fully determined by experimental measurements and consistent with calculations given in \cite{safronova2011blackbody}.  It gives $\Delta\alpha_0(\omega_{1068})=-15.66(16)+0.12(5)\,\mathrm{a.u.}$ with $0.12(5)\,\mathrm{a.u.}$ giving the estimated correction from the uv terms, which is less than the uncertainty from the experimentally determined component.  At this level of inaccuracy attention should be given to the dependence on the measurement wavelength, which is $-0.186\,\mathrm{a.u./nm}$ at the 1068\,nm used in \cite{wei2024improved}.  The experimentally determined extrapolation is a factor of three more accurate than the value of $-14.93(44)\,\mathrm{a.u.}$ used in \cite{wei2024improved}, which was derived from theory.  The difference of 0.60(46)\,a.u. is in fair agreement given that their extrapolation is based on theory and their uncertainty does not correctly consider correlation in the resulting terms of their theoretical treatment \cite{barrett2025extrapolation}.

It should also be noted that Eq.~\ref{CaDSP} will accurately determine $\Delta\alpha_0(\omega)$ at shorter wavelengths between the resonances at $\omega_{854}$ and $\omega_{397}$.  Consequently a similar comparison measurement with Al$^+$ could be also be done at e.g. $650$ or $780\,\mathrm{nm}$ for which inexpensive high-power lasers are readily available.  The additional measurement would provide an accurate, fully experimental determination of $\Delta\alpha_0(0)$ for Al$^+$ without the need for theoretical extrapolations.

\section{Summary}
In summary, a zero-crossing of the Ba$^+$ differential scalar polarizability has been located at 623.603\,13(17)\,THz.  This measurement has allowed us to infer a ratio of reduced matrix elements $\redmel{P_{3/2}}{r}{S_{1/2}} /\redmel{P_{1/2}} {r}{S_{1/2}}=1.411\,81(13)$, which provides a stringent test of atomic structure calculations and has good agreement with the value of 1.4116(2) reported in \cite{roberts2023electric,cserveny2025theoretical}.  Our value is within $1.8\sigma$ of the ratio reported in \cite{woods2010dipole} with an order of magnitude improvement in the precision. Using this ratio, the branching fraction reported in \cite{zhang2020branching}, the zero-crossing reported in \cite{chanu2020magic}, and an updated value of $\redmel{P_{1/2}} {r}{S_{1/2}}$ given by Eq.~\ref{Eq:MErevised}, we have given a parameterization of $\Delta\alpha_0(\omega)$ that has a fractional inaccuracy $\leq 0.23\%$ for frequencies up to 450\,THz.  This is at least a factor of ten more accurate than can be reasonably determined from measurements using calibrated beam parameters and will enable accurate calibration of polarizabilities for other clock transitions via comparison with Ba$^+$ as discussed in \cite{barrett2019polarizability}.  In particular, it will enable more than an order of magnitude improvement in the uncertainty of $\Delta\alpha_0(0)$ for the clock transitions in Lu$^+$, which is currently the leading systematic for the $^1S_0-{}^3D_1$ clock transition, giving access to clock inaccuracies below $5\times 10^{-20}$ for this system.  A further independent check on the revised matrix $\redmel{P_{1/2}} {r}{S_{1/2}}$ given here should be possible using the measurement technique demonstrated for Ca$^+$ \cite{hettrich2015measurement}.

The methodology we have used to characterize polarizability curves of a wide frequency range is applicable to other earth-alkaline ions, notably Ca$^+$, Sr$^+$, and Ra$^+$.  In the case of Ca$^+$, we have shown that our approach allows extrapolation from an accurate determination of $\Delta \alpha_0(0)$, which is directly relevant to the results reported in \cite{wei2024improved}.  In this case we have assumed a matrix element ratio of $\sqrt{2}$.  However, the same use of a zero crossing could be used to confirm this.  Despite the shorter wavelength of transitions to $P_{1/2}$ or $P_{3/2}$, the relatively small fine-structure splitting means the zero crossing is more heavily determined by these two transitions than is the case in Ba$^+$.    

\acknowledgements
This project is supported by the National Research Foundation, Singapore through the National Quantum Office, hosted in A*STAR, under its National Quantum Engineering Programme 3.0 Funding Initiative (W25Q3D0007) and under its Centre for Quantum Technologies Funding Initiative (S24Q2d0009).

%

\end{document}